\documentclass{aa520}
\usepackage{txfonts}
\usepackage{graphicx}
\usepackage[authoryear]{natbib}
\providecommand{\LyX}{L\kern-.1667em\lower.25em\hbox{Y}\kern-.125emX\@}

\newcommand{\Msun}{\mbox{$M_\odot$}}
\newcommand{\Teff}{\mbox{$T_{\rm eff}$}}
\newcommand{\Mbol}{\mbox{$M_{\rm bol}$}}
\newcommand{\simgt}{\lower.5ex\hbox{$\; \buildrel > \over \sim \;$}}
\newcommand{\simlt}{\lower.5ex\hbox{$\; \buildrel < \over \sim \;$}}
\def\aap{\ref@jnl{A\&A}}
\def\aj{{AJ}}%
\def\apj{{ApJ}}%
\def\apjl{{ApJ}}%
\def\apjs{{ApJS}}%
\def\aap{{A\&A}}%
\def\aaps{{A\&AS}}%
\def\memras{\ref@jnl{MmRAS}}%
\def\mnras{{MNRAS}}%

\begin{document}

\title{Lithium during the AGB evolution in young open clusters of the Large
Magellanic Cloud 
\thanks{based on observations collected at the European Southern Observatory, 
La Silla, Chile}}


\author{C. Maceroni \inst{1}, V. Testa \inst{1}, B. Plez \inst{2}, P. Garc\'\i a
Lario \inst{3} \and F. D'Antona \inst{1}
}

\offprints{C. Maceroni,\\
\email{maceroni@coma.mporzio.astro.it} }

\institute{INAF, Osservatorio Astronomico di Roma , via Frascati 33, I-00040
Monteporzio C. (RM), Italy \\
 \email{surname@coma.mporzio.astro.it} 
\and GRAAL, Universit\'e Montpellier II, F-34095 Montpellier cedex 5, France
\and ISO Data Centre, Science Operations and Data Systems Division, Research and 
Scientific Support Department of ESA, VILSPA, 28080 Madrid, Spain
}

\date{Received ; accepted }

\authorrunning{Maceroni et al.}

\titlerunning{Lithium production in LMC young open clusters}

\abstract {
We present the results of mid-resolution spectroscopy in the \ion{Li}{I} 6708 \AA~
spectral region of Asymptotic Giant Branch (AGB) stars belonging to young
open clusters of the Large Magellanic Cloud. Most stars belong to the
clusters \object{NGC 1866} and \object{NGC 2031}, which have an age of $\simeq 150$~Myr.
Lithium lines of different strength are detected in the spectra of stars
evolving along the AGB, in agreement with theoretical predictions. According
to stellar evolutionary models, at the start of the AGB the stars should all
show a low residual lithium abundance as a consequence of  dilution during
the previous evolutionary phases. The most luminous and cooler thermally
pulsating  AGB stars, if they are massive enough, once in the AGB go first
through a phase of Li destruction, which is followed by a phase of strong
lithium production and further destruction.
The production of lithium on the AGB is in particular  explained by the onset
of the {}``Hot Bottom Burning\char`\"{} (HBB) process. Our most conclusive
results are obtained for the populous cluster \object{NGC 1866} where: the
`early--AGB' stars show a weak Li line, which can be attributed to the
dilution of the  initial abundance; one of the two luminous stars seem to
have completely depleted lithium, as no line is detected;  the second one
shows a deep lithium line, whose strength can be explained by lithium
production. The bolometric magnitude of these stars are consistent with the
values predicted by the theory, for the mass evolving on the AGB of \object{NGC 1866},
at which lithium is first destroyed and then produced (\Mbol $\simeq-6$).
We  also analyze the infrared luminosities (ISOCAM data) of these stars, to
discuss if their evolutionary phase precedes or follows the lithium
production stage.
More intriguing and less clear results are obtained for the most luminous
stars in \object{NGC 2031}: the brightest star seems as well to have destroyed
lithium, while the second one shows a strong lithium line. However its
bolometric luminosity --derived from the near--IR photometry, is much lower
(\Mbol $\simeq -5.2 \pm 0.15$)  than that expected from HBB models.
Although low luminosity lithium rich AGB stars are also known, whose appearance
is attributed to non--canonical mixing processes, it is not clear why two
almost coeval clusters show such a different behaviour.
It is also possible that this star does not belong to \object{NGC 2031}.
Finally we suggest the observational tests that could shed further light on
this matter.}

\maketitle


\section{Introduction}
The evolution of lithium in intermediate mass stars is a powerful probe of
the convective envelope conditions. In particular, during the Asymptotic
Giant Branch (AGB) phase in stars of mass large enough for the Hot Bottom
Burning (HBB) process to occur, the lithium abundance has a complex
behaviour, whose knowledge can shed light on the efficiency of convection and
on the occurrence of convective overshooting in the preceding phases. The
most suitable targets for such a study are the relatively massive and
thermally pulsating (TP) AGB stars in (young) clusters, whose mass
can, in principle, be derived by fitting the cluster color-magnitude diagram.
The need for a relatively well populated AGB naturally favors the choice of
young clusters in the Large Magellanic Cloud.

We carried out, therefore, an observing program to explore the lithium
abundance in the TP--AGB phase of stars  belonging to four Large Magellanic
Cloud open clusters and whose initial mass was determined by fitting the
morphology of the color-magnitude diagram (CMD), turnoff included. We
obtained mid-resolution spectra of several stars in \object{NGC 1866} and \object{NGC 2031}
(which have ages \( \simeq 150 \)~Myr), of one star in the younger (\( \simeq
100 \)~Myr old) cluster \object{NGC 2214} and of one star in the older \object{NGC 2107} (\(
\simeq 250 \)~Myr) \citep[e.g.][]{corsi94,girardi95}. The most
luminous (three) stars in \object{NGC 1866} and (two) in \object{NGC 2031}, and the stars in the
other clusters, were selected from the list of \citet[][ hereafter FMB90]{frogel90}. 
Being these also the latest spectral type clusters' stars, they
are good TP--AGBs candidates. Additional stars were selected in \object{NGC 1866} and
\object{NGC 2031}, as good candidates for the `early--AGB' phase of evolution on the
base of our own near IR photometry.

We looked for and derived the strength of the lithium line at \( \lambda
= 6707.8 \) \AA, and explored its dependence on the AGB luminosity and on the
cluster age. The observed spectra were compared with synthetic ones, for  
evaluation of the lithium abundance. Though a higher dispersion is
required for a precise abundance determination, our mid-resolution spectra
already provide interesting results.

In the next Sections we present the theoretical background of our project
(Sect. 2), the criteria for cluster and target selection (Sect. 3), the
observations and data reduction (Sect. 4), the analysis of the lithium
abundance (Sect. 5)  and its theoretical implications (Sect. 6), the analysis
of the evolutionary stage represented by the selected stars (Sect. 7), and the
final discussion and conclusions (Sect. 8).

\section{Theoretical background}

The term `Hot Bottom Burning' (HBB) refers to nuclear processing that can
take place at the bottom of the convective envelope of massive Asymptotic
Giant Branch (AGB) stars during the TP phase. The observations of these stars
in the Magellanic Cloud fields  \citep{smith89,smith90,plez93,smith95}
reveal that the most luminous AGB stars are lithium rich. 
That is indeed satisfactorily explained by the
theoretical models including HBB.

\begin{figure}
\centerline{\resizebox{8.8cm}{!}{\rotatebox{0}{\includegraphics{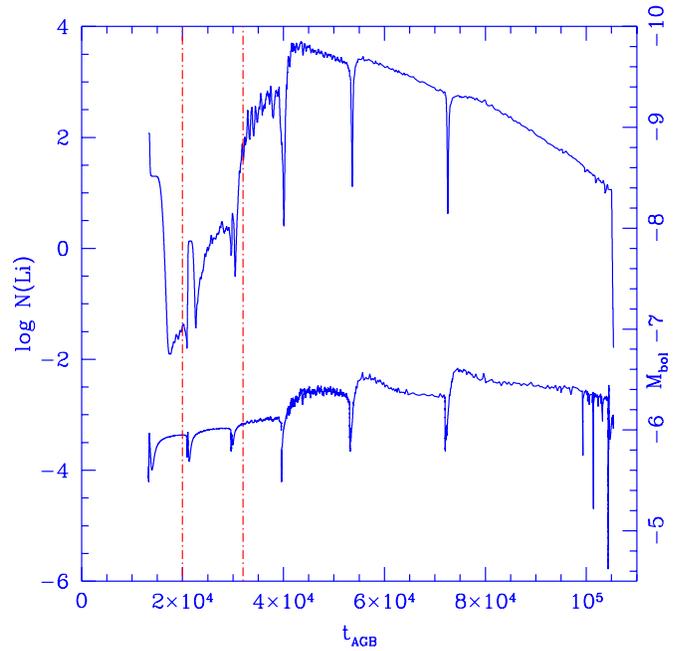}}}}
\caption[]{ The evolution of the bolometric magnitude (bottom curve, scale on
the right) and of the lithium abundance (top curve, scale on the left) during
the TP phase for a 4 \Msun  with Z=0.01 \citep{ventura99}. The
vertical lines indicate the possible evolutionary phase of the two most
luminous AGB stars in  \object{NGC 1866}.
\label{fig1} }
\end{figure}

During the giant phase preceding the AGB, lithium is depleted by convective
dilution. Therefore, if we find on the TP-AGB an abundance much higher than
the residual from the preceding phases, we can safely conclude that lithium
is manufactured `in situ'. Both the computations performed by considering only
envelope models \citep{sackmann74} and those of complete evolutionary
models \citep{sackmann92} show that a temporary enhancement of
the lithium abundance in the atmosphere --- in agreement with the previously
mentioned observations --- can be achieved through the chain \(
\element[][3]{He}+\element[][4]{He} \rightarrow \element[][7]{Be} \rightarrow \element[][7]{Li} \),
provided that the mixing is a non-instantaneous process \citep{cameron71}.
The occurrence of nuclear burning at the bottom of the convective
envelope is exemplified in Fig. \ref{fig1}, which shows the time evolution of
luminosity and lithium abundance for a 4\Msun\ stellar model ascending the
AGB \citep[from][]{ventura99}. The model has helium and metal mass fraction,
respectively, Y=0.26 and Z=0.01.
 
The lithium evolution in the AGB can be sketched as a sequence of four
different phases:
 i) at the beginning of the AGB, the stellar photosphere has a residual
lithium content that is a remnant from a previous convective dilution phase;
 ii) lithium is totally destroyed when the temperature at the
bottom of the envelope increases above 10\( ^{7} \)~K; iii) later on,
production via the Cameron -- Fowler mechanism proceeds and a high lithium
abundance is reached; iv) finally, after \( \simeq 10^{5} \)
 yr, lithium is
depleted again, as a consequence of the exhaustion  of \( ^{3} \)He in
the convective envelope. The stellar mass is,
 at this stage, 1.7 \Msun, its
mass loss rate is \( \simeq 2 \times 10^{-5} \) \Msun yr$^{\mathrm{-1}}$, and the carbon
oxygen core mass is 0.82 \Msun. The evolution will last about 5 \( \times 10
^{4} \) yr longer, until the whole envelope is lost and the star evolves to
the white dwarf stage.

The determination of $^{7}$Li abundance along the AGB
provides information about the physical processes influencing the
efficiency and the extent of mixing, and can help to validate the
physical modeling of these processes in stellar interiors. 

\section{Clusters and  target selection}

The aim of our observations was to understand the correlation between the
mass of the stars evolving along the AGB and the lithium abundance. This
cannot be done on field giants, because their masses are unknown,
while the age of clusters (and therefore the masses of the stars evolving
along the AGB)  can be derived from the turnoff and/or from the location of
the Helium core burning (clump) stars. We  mainly used for this purpose the
recent models by Ventura et al., in preparation, described in \citet{kalirai01}.
For the same models, the evolution is completed through the HBB phases is
described in \citet{ventura00}.

\begin{figure}
\centerline{\resizebox{8.8cm}{!}{\rotatebox{0}{\includegraphics{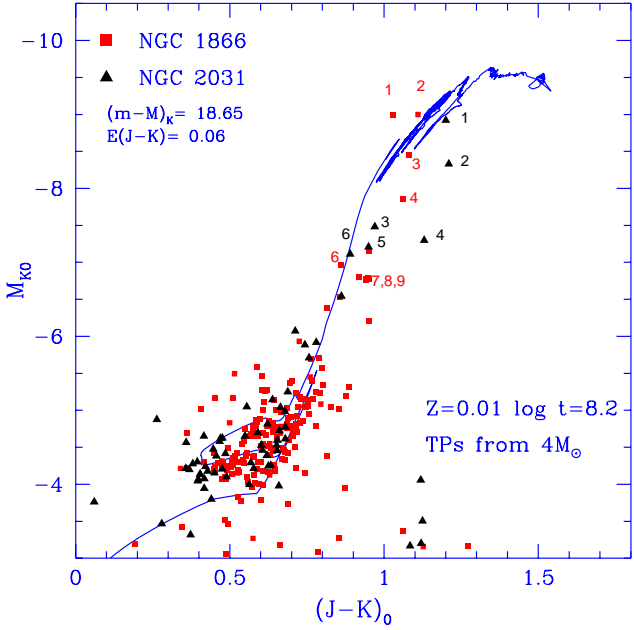}}}}
\caption[]{De-reddened CM diagram of  NGC 1866 (squares) and NGC 2031
(triangles).
The target stars are identified according to the numbering of Table
\ref{liststars}. An isochrone is also shown (from Ventura et al., in prep.)
corresponding to a mass in AGB of $\sim$ 4.3 \Msun,   together with the TP
AGB evolution of a 4 \Msun\ star (the difference in mass with  respect to the
isochrone is of no relevance here). The transformation from the theoretical
to the observational diagram was performed by means of the relations of
BCP98.
\label{fig2}
}
\end{figure}

Unfortunately there are not many AGB stars in the young LMC open clusters
and it is not well known which ones, among the few M type giants in the
clusters, are in the TP phase. Our sample of targets was assembled including
the candidates from the list of FMB90  adding a few objects
in the clusters \object{NGC 1866} and \object{NGC 2031} that, from our own near IR photometry
(Testa, unpublished), come out to be `early AGB' candidates (see Fig.
\ref{fig2}). 

\subsection{Cluster selection}
\subsubsection{\object{NGC 1866}}

This cluster, one of the most populous of the LMC, has been very much debated
in the literature, as it turns out to be a good test of the possible
occurrence of core overshooting during hydrogen burning \citep[see the latest
discussion by][]{barmina02}. \citet{testa99} carefully compared
the optical data with models, deriving ages which vary according to the input
evolutionary tracks. The authors used either `standard' models, or models
with an `extended core' during the hydrogen burning stage, that is, an
artificially enlarged convective core (without a physical reason for this
extension). This simplified treatment provides limits for the evolving masses
which range from \( \sim 4.4-5\)~\Msun\ for the standard case,  to 3.9~\Msun\
for the extended core computations (A. Chieffi, private communication).
We can compare these results with the models by Ventura et al., in preparation.
These include a more physical description of core overshooting, by considering
non--instantaneous mixing in the whole core, and extending it beyond the
formal convective region. The  mixing velocities of these models are taken from 
the convective model of \citet{canuto96}. In the overshooting region, the
velocities are extrapolated by assuming an exponential decay (consistent with
large eddy numerical simulations of convection). The decay scale length is
calibrated by means of several observational parameters (first of all, the
main sequence observational width), through a parameter \( \zeta  \) (the 
larger the value of $\zeta$, the slower the decay). Models
with  \( \zeta =0.02-0.03 \)\  reproduce the observations \citep{ventura98}. 
The same overshooting is also used for the helium--core
burning phases. These new models roughly confirm the \citet{testa99}
computations. The turnoff location and the luminosity level of the red clump
stars lead us to date the cluster at ages \(\log \mathrm{t} = 8.05 \div 8.25\):
the smaller value refers to models without overshooting  (\( \zeta =0 \)), 
the larger one to those with the largest overshooting  (\( \zeta =0.03 \)), 
and to a composition \(Y=0.26\) and \(Z = 0.006 \div 0.01 \).
 The corresponding evolving mass in the AGB ranges from 4.3 (\( \zeta =0.03 \))
 to 5\,\Msun\ (\( \zeta = 0 \)).

It has to be stressed that the great difference in the evolving masses
between scenarios with and without overshooting does not correspond to  a
large difference in the AGB evolution, as the carbon oxygen core
mass, that mainly determines the AGB evolution, is about the same in both
cases. Both the 4.3 \Msun\ ($\zeta =0.03$) and the 5 \Msun\ ($
\zeta =0 $) models with Z=0.01 will evolve through a HBB phase and produce
lithium, at least if the Full Spectrum of Turbulence (FST) efficient
convection model \citet{ventura00} is adopted.

\subsubsection{\object{NGC 2031}}

The CMD of this cluster looks very much similar to that of \object{NGC 1866}, although
it is much less populated. Because of its location, in a highly populated 
region 
of the LMC, the analysis of the photometric data is rather difficult. This,
together with the absence of dynamical studies, leaves open the question of 
the cluster membership of some of its stars. 
\citet{mould93} estimated ages of 140\( \pm\)20~Myr by means of an 
isochrone fitting to the optical CMD, and stressed
the similarity with \object{NGC 1866}. 

\begin{table*}
\caption[]{Photometric and spectroscopic observations}
\label{liststars}
$$
\begin{array}{cccccccccl}
\hline
\noalign{\smallskip}
 Star  &  J_0  & H_0  & K_0   & R_0 & V_0 & B_0 &  N_{obs} &  S/N  & \overline{V}_{helio} \\
\noalign{\smallskip}
\hline
\noalign{\smallskip}	  	  		   		   	  	
N1866\#1 &  10.70 &   9.89  &  9.67   & 13.81  & 15.38  & 17.06 &   3  &42, 37, 40 &	305 \pm 5	\\ 
N1866\#2 &  10.77 &   9.91  &  9.66   & 13.80  & 14.97  & 16.61 &   2  &  48, 55   &	304 \pm 5	\\ 
N1866\#3 &  11.29 &  10.44  & 10.21   & 14.14  & 15.67  & 17.24  &   2 &  39, 33   &	306 \pm 6	\\ 
	 &	&	   &			&        &         &         &        &        &        \\
N1866\#4 &  11.87 &  11.01  & 10.81   & 13.72  & 14.84  & 16.52  &   1 &  42	   &	307 \  pm 2.\\ 
N1866\#5 &  12.45 &  11.66  & 11.50   & 13.82  & 14.86  & 16.43  &   1 &  65	   &	298 \pm 1	\\ 
N1866\#6 &  12.56 &  11.79  & 11.70   & 13.85  & 14.75  & 16.21  &   1 &  57	   &	298 \pm 3.4	\\ 
N1866\#7 &  12.78 &  12.03  & 11.86   & 14.11  & 14.91  & 16.34  &   1 &  52	   &	299 \pm 2.7	\\ 
N1866\#8 &  12.83 &  12.04  & 11.88   & 14.21  & 15.23  & 16.74  &   1 &  30	   &	300 \pm 2.5	\\ 
N1866\#9 &  12.83 &  12.07  & 11.89   & 14.12  & 15.14  & 16.58  &   1 &  12	   &	307 \pm 7	\\ 
 & & & & & & & & & \\
N2031\#1 &  10.93  & 10.12  &  9.73  &  14.61  & 		& 		 &   2 &  52, 60   &	235 \pm 6   \\ 
N2031\#2 &  11.53  & 10.62  & 10.32  &  13.66  &		&		 &   2 &  21, 22   & 	259 \pm 6   \\ 
N2031\#3 &  12.14  & 11.33  & 11.17  &  13.54  &		&		 &   2 &  53, 66   &	248 \pm 2.2 \\ 
N2031\#4 &  12.48  & 11.64  & 11.35  &  15.08  &		&		 &   0 &   -	   &	   -	    \\ 
N2031\#5 &  12.39  & 11.59  & 11.44  &  13.81  &		&		 &   1 & 57		   &	225 \pm 2   \\ 
N2031\#6 &  12.43  & 11.64  & 11.54  &  14.91  &		&		 &   1 & 40		   &	237 \pm 8   \\ 
 & & & & & & & & & \\
N2214\#1^\mathrm{a} &  10.88  & 10.12  &  9.90  &   	   & 	   &		 &   1 & 56		   &	260 \pm 3   \\
 & & & & & & & & & \\
N2107\#1^\mathrm{a} &  10.91  & 10.07  &  9.79  &		   &	   &		 &   1 & 42		   &	272 \pm 7   \\
\hline
\end{array}
$$
\begin{list}{}{}
\item[$^{\mathrm{a}}$] The magnitudes of \object{NGC 2214} and \object{NGC 2107} are not de-reddened. 
\end{list}
\end{table*}

\subsubsection{\object{NGC 2214} and \object{NGC 2107}: two comparison clusters}

These two clusters have been selected for comparison with the \object{NGC 1866} and
\object{NGC 2031}, the two main clusters analyzed in this paper.
 The first one is definitely younger as can be seen by looking at the
CMD published by \citet{banks95} or at the one previously available from
\citet{robertson74}. The age estimate for \object{NGC 2214} ranges from 32~Myr 
\citep{elson91} to \( \sim  \) 100~Myr \citep{girardi95}. 
We have re-calibrated the cluster age by using synthetic CMDs built with a grid
 of models from \citet{ventura00}, obtaining \(t  \sim  100\)~Myr. \object{NGC 2107} is, on the other
hand, older than \object{NGC 1866} and \object{NGC 2031}, as found by 
\citet{corsi94} with a (V, B$-$V) CMD. 
\citet{girardi95} suggest an age of 250~Myr for this
cluster. Our re-calibration, by using the same method applied  for \object{NGC 2214},
is in agreement with the above value.

\subsection{Target selection}

The final sample of target stars is listed in Table \ref{liststars}. Each
star is identified with the cluster name and is labeled in order of
increasing K magnitude.  Table \ref{liststars} lists the magnitudes in
various bands. The B, V photometry of \object{NGC 1866} is from \citet{testa99};
the R photometry for all clusters but \object{NGC 2107} was obtained from calibration
shots taken in November 1999 with SUSI2 - NTT; the J, H, K magnitudes of NGC
1866 and 2031 were obtained in 1995 at ESO 2.2m - IRAC2.   However, for the
three brightest stars of \object{NGC 1866} we had to adopt the   J, H, K magnitudes by
FMB90. This choice was preferred because these bright objects fall in the
non-linear regime of the infrared IRAC2 detector. The B and V magnitudes of
\object{NGC 2214} are from the original work of \citet{robertson74}.

In order to de-redden the magnitudes of \object{NGC 1866} we adopted $A_J=0.09$, 
$A_H=0.06$, $A_K=0.03$ and  $A_V=0.25$. The absorption in the R and B bands 
were derived
by means of the \citet{rieke85} relations for absorption in
different bands. The reddening of \object{NGC 2031} is more uncertain: 
\citet{mould93} gives E(B$-$V) spanning from 0.06 to 0.18. On the basis of the good
match of the red clump of the two clusters we decided to adopt the same
reddening values for both of them.

Fig. \ref{fig2} shows the near IR color magnitude diagram for \object{NGC 1866} and NGC
2031 (Testa, unpublished), from which most of the targets have been selected,
and a superposed  isochrone by Ventura et al., in preparation,  describing the 
clump
stars and the early AGB evolution. An  age of log t=8.2 was chosen on the
basis of the fit of the whole CM diagram \citep[V versus B$-$V from][]{testa99}, including the turnoff.

The TP AGB phase is sketched on the same plot, by adopting an evolutionary
mass of 4\Msun, (Y=0.26, Z=0.01) from \citet{ventura00}. The track
includes mass loss according to \citet{blocker95} formulation, with
the Reimer's parameter $\eta$\  fixed at 0.01.

Only one star per cluster could be selected in  \object{NGC 2214} and \object{NGC 2107}
from the sample of FMB90, namely star B69 of \object{NGC 2214} 
\citep[the notation refers to the work of][]{robertson74}, and star no. 6 of \object{NGC 2107}.

\section{Observations and data reduction}

The target stars were observed during two nights, in December 1999,
with the ESO New Technology Telescope (NTT) and EMMI in the red mid-resolution
spectrographic mode. The selected grating (\#4)  has a nominal resolution
of 5500 (for the 1\char`\"{}-wide slit we used), however the seeing conditions
of one night were so good (0.5\char`\"{}--0.6\char`\"{}) that for a
part of the spectra the actual resolution is actually higher than the
nominal value. 
A total of 23 long slit spectra were obtained for 16  out of the 17 stars in 
the initial sample,  all of them are 640 \AA\ long and are centered on the 
\ion{Li}{I} 6707.8\ \AA\ line; the details are given in the last column of Table 
\ref{liststars}.
Some of the stars of \object{NGC 1866} and \object{NGC 2031} were observed several times. The 
number of spectra available for each target is listed in Column 7 of Table 
\ref{liststars} together with the mean S/N ratio of the individual spectra.
Note that star \#4 of \object{NGC 2031}, though present in the initial target list,
was not observed because of its rather low priority on the basis
of its blue color and uncertain classification, however to avoid confusion
with the data file names we kept the initial numbering of the sample. 
The data reduction was performed by means of standard
techniques (the long slit reduction routines of the IRAF \footnote{IRAF is distributed by the National Optical Astronomical Observatories, which
are operated by the Association of the Universities for Research in Astronomy, Inc., under cooperative agreement with the National Science Foundation} package).
The steps were bias subtraction, flat fielding, (partial) cosmic ray
removal, normalization, wavelength calibration.
The Doppler shifts were derived from H\( \alpha  \) and a few other
reference lines (mainly \ion{Fe}{I}), and the corresponding radial velocities
(or a mean value when more  spectra were available) are given in the last 
column of Table \ref{liststars}. This simple derivation provides relatively
large errors, but the results are at least consistent with those from more 
accurate estimations: for instance, the Doppler shifts derived for the whole 
sample of nine \object{NGC 1866} AGB stars,  yield a mean velocity 
\( V=302.7 \pm 3.6 \) km s$^{-1}$, 
in good agreement with previous dynamic studies: according to 
\citet{fischer92} the systemic velocity of \object{NGC 1866}  
is \( 301\pm 1.2 \) km s$^{-1}$. The small deviations of the individual stars 
from the mean value confirm as well their cluster membership. 

\begin{figure}
\centerline{\resizebox{8.8cm}{!}{\rotatebox{0}{\includegraphics{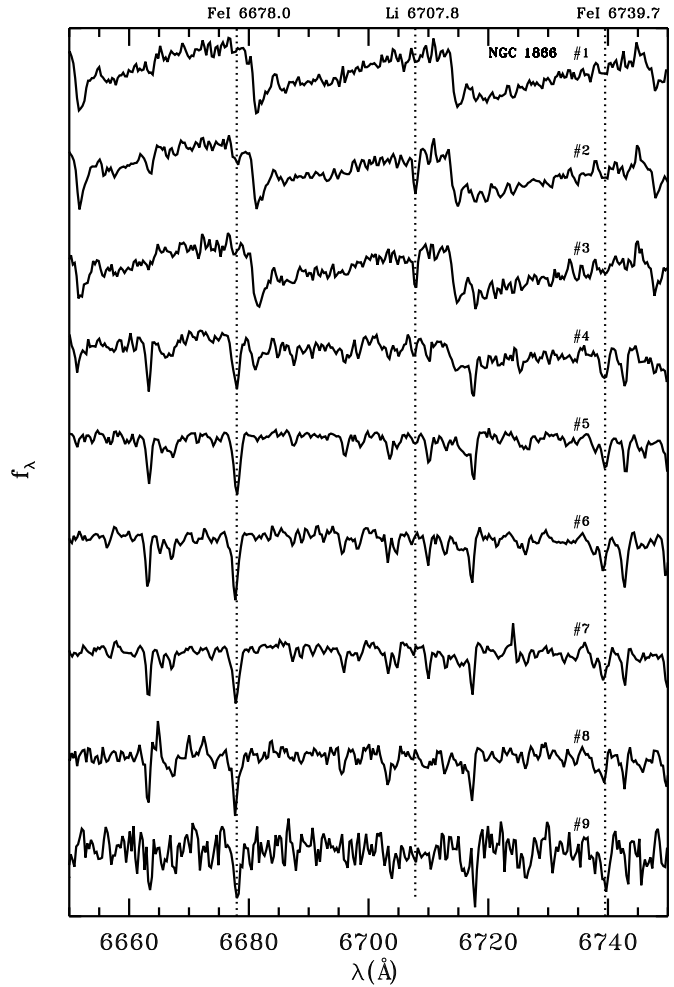}}}}
\caption[]{The spectra in the \ion{Li}{I}  6707.8\ \AA\ region of the nine AGB stars observed in \object{NGC 1866}. The stars are ordered from top to bottom by decreasing near-IR luminosity.
\label{fig3}}
\end{figure}

\begin{figure}
\centerline{\resizebox{8.8cm}{!}{\rotatebox{0}{\includegraphics{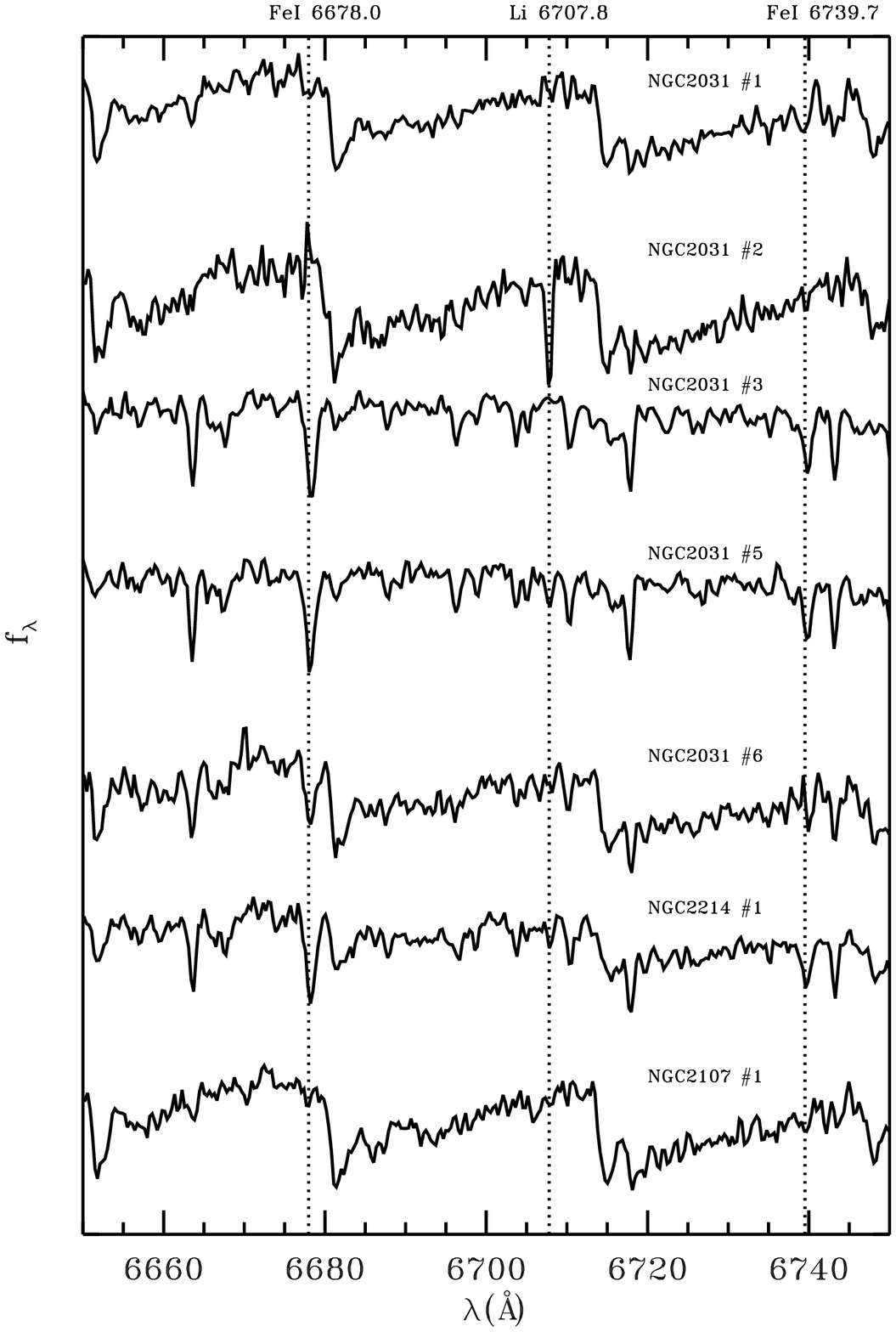}}}}
\caption[]{The spectra in the  \ion{Li}{I} 6707.8\ \AA\ region of the seven AGB stars observed in \object{NGC 2031}, together with those of the  AGB stars observed in the comparison clusters \object{NGC 2214} and  \object{NGC 2107}.
\label{fig4}}
\end{figure}

Figs. \ref{fig3} and \ref{fig4} show the reduced spectra normalized to the 
maximum flux  in the wavelength range displayed.

\begin{figure}
\centerline{\resizebox{8.8cm}{!}{\rotatebox{0}{\includegraphics{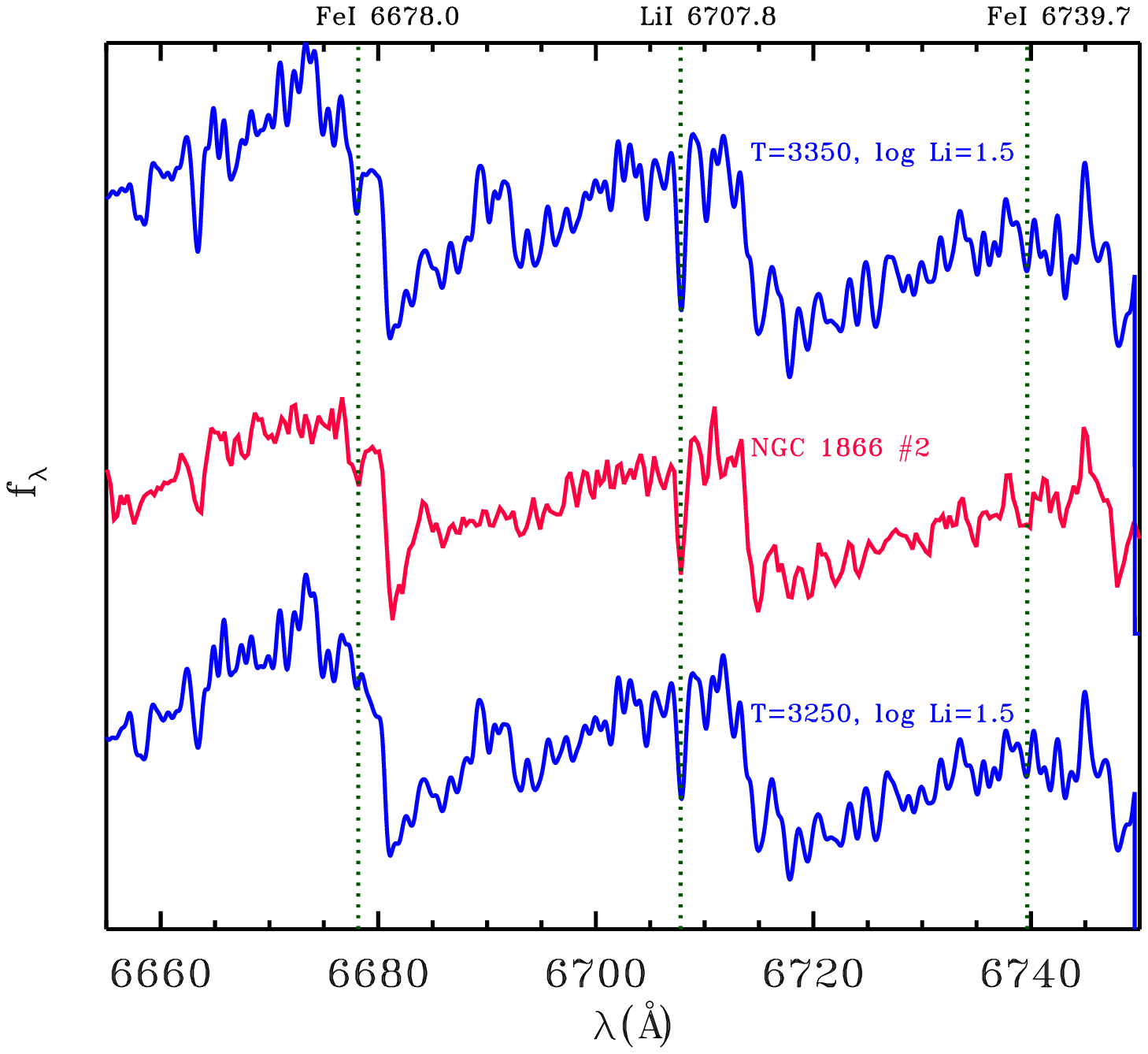}}}}
\caption[]{Comparison among the spectrum of NGC 1866\#2 and two synthetic
spectra computed for \Teff\ = 3250, 3350 K, $\log N(\element[][]{Li})=1.5$,
$\log g=0.0$ and [\element[][]{Fe}/\element[][]{H}] $=-0.4$
\label{fig5}
}
\end{figure}

A \ion{Li}{I} line at 6708 \AA\ of decreasing
strength is clearly visible in stars \#2, \#3, \#4 of \object{NGC 1866}, but
is absent in the K-band brightest star \#1. The same seems to happen in
\object{NGC 2031}, where again the spectrum of star \#2 shows a strong \ion{Li}{I} line, 
while that of star\#1 one does not.  Fig. \ref{fig5} shows a comparison
between the observed spectrum of \object{NGC 1866} and two synthetic spectra with 
reasonable \Teff\ and lithium content, where it is evident that the 
lithium abundance of this star must be large. A detailed analysis of the 
lithium abundances of each star is presented in the next Section.

In addition, three different images of the \object{NGC 1866} field obtained
with ISOCAM, the mid-IR camera onboard the Infrared Space Observatory (ISO),
were retrieved from the ISO Data Archive (TDT numbers: 59\,000\,438, 59\,000\,439 
and 59\,000\,440) in order to study the mid-IR emission
of the AGB sources under analysis in this paper 
 and search for other possible heavily obscured AGB stars in the same field 
not listed in Table \ref{liststars}. 
The images were taken through the broadband 
ISOCAM LW1, LW2 and LW10 filters, whose passbands are centered at 4.5, 6.7 
and 12.0 $\mu$m respectively \citep{blommaert02}.
 They were originally part of the proposal  TTANABE.REDSTAR. They were all taken on 
 28 June 1997 and cover a region  of 192 \char`\"{} $\times$ 192 \char`\"{} with a pixel scale 
 of 3.0 \char`\"{} pix$^\mathrm{-1}$.

 Data reduction was performed using the  {}``CAM Interactive Analysis \char`\"{}  (CIA, version 4.0) 
 starting from the raw data products which were corrected for dark current, glitches,
transients and flat-field,  following the standard routines available within 
CIA. An accuracy better than 20\% is expected even at these very low flux
levels in the absolute flux assigned to the end products.

\section{Determination of the lithium abundances}

The analysis was performed by comparing our observations with a grid of
synthetic spectra with different temperatures and lithium abundances.

 The synthetic spectra were calculated with the Turbospectrum 
program \citep[]{alvarez98}, using MARCS model atmospheres \citep{gustafsson75,  
plez92, asplund97, gustafsson03}. The models and the 
spectra are spherically symmetric, at LTE, and include up-to-date continuous and line 
opacities  for a large number of atoms and molecules. Some details are 
provided in \citet{hill02}. The line lists for VO, ZrO, CN and 
TiO (Plez 1998), and atomic lines \citep[VALD, ][]{kupka99}
were used. The spectra were initially computed at a resolution of
0.01\AA, and then degraded to match the resolution of the 
observations.

The stellar parameters needed for synthetic spectra computation (\Teff, 
$\log g$) were initially derived from the photometry.
The K band bolometric correction \( BC_{K} \)\ was obtained  by using the \(
BC_{K} \)\ versus \( V-K \)\ relation by \citep[][ BCP98]{bessell98} for the
nine stars of \object{NGC 1866} that have \( V-K \)\ colors. The same stars were used
to derive a  \( BC_{K} \)\ versus \( J-K \)\ relation that was needed to
derive the bolometric correction for the stars of the other clusters,  for
which \( V-K \)\ is not available.

\begin{table*}
\caption[]{Observed and derived target properties}
\label{magn}
$$
\begin{array}{p{0.1\linewidth}clccrclcp{0.18\linewidth}}
\hline
\noalign{\smallskip}
ID & M_k & BC_k & \Mbol & \log \Teff & \log g & \log \Teff & \log g & \log N(\mathrm{Li}) & Comment\\
& & & &  \multicolumn{2}{c}{\mathrm{from \: phot.}} & \multicolumn{2}{c}{\mathrm{adopt.\: values}} & & \\
\noalign{\smallskip}
\hline
& & & & & & & & & \\
\noalign{\smallskip} 
N1866\#1 &  -8.93 &  2.94 & -5.99  &  3697. &   0.01 & 3350 & 0.0 & <-0.5   &  ISO detected\\
N1866\#2 &  -8.94 &  2.90 & -6.04  &  3529. &  -0.09 & 3250 & 0.0 & 1.5   \pm 0.5 & ISO detected \\
N1866\#3 &  -8.39 &  2.91 & -5.48  &  3592. &   0.16 & 3450 & 0.2 & 0.5   \pm 0.5  & ISO detected \\
N1866\#4 &  -7.79 &  2.68 & -5.11  &  3634. &   0.33 & 3750 & 0.5 & -0.25 \pm 0.3  & \\
N1866\#5 &  -7.10 &  2.52 & -4.58  &  3866. &   0.65 & 3950 & 0.8 & -0.25 \pm 0.3  & \\
N1866\#6 &  -6.90 &  2.45 & -4.45  &  4055. &   0.78 & 4100 & 0.9 & +0.0  \pm 0.3  & \\
N1866\#7 &  -6.74 &  2.45 & -4.29  &  3929. &   0.79 & 4000 & 0.9 & -0.25 \pm 0.3  & \\
N1866\#8 &  -6.72 &  2.52 & -4.20  &  3866. &   0.80 & 3950 & 0.8 & +0.0  \pm 0.25 & \\
N1866\#9 &  -6.71 &  2.50 & -4.21  &  3887. &   0.81 & 4000 & 0.9 & <0.5 (?)  & very noisy  \\
\noalign{\smallskip}                     
\hline                                     
\noalign{\smallskip}                       
N2031\#1 &  -8.87 &  3.06 & -5.81  &  3340. &  -0.10 &  3450 & 0.0 &<-0.5     &      \\
N2031\#2 &  -8.28 &  3.08 & -5.20  &  3319. &   0.14 &  3350 & 0.0 &  2.5 \pm 0.5  & \\
N2031\#3 &  -7.43 &  2.59 & -4.84  &  3824. &   0.53 &  3950 & 0.8 & <-0.5    &      \\
N2031\#4 &  -7.25 &  2.92 & -4.33  &  3487. &   0.57 &       &     &          & not observed           \\
N2031\#5 &  -7.16 &  2.55 & -4.61  &  3866. &   0.64 &  3950 & 0.8 & 0.2  \pm 0.2  & \\
N2031\#6 &  -7.06 &  2.42 & -4.64  &  3992. &   0.68 &  3450 & 0.2 &<-0.5     &      \\
\noalign{\smallskip}                       
\hline                                     
\noalign{\smallskip}                       
N2214\#1 &  -8.70 &  2.61 & -6.09  &  3802. &   0.02 &  3650 & 0.3 & -0.5 \pm 0.3 & \\
\noalign{\smallskip}                  
\hline                                     
\noalign{\smallskip}                        
N2107\#1 &  -8.81 &  2.90 & -5.91  &  3508. &  -0.05 &  3450 & 0.2 & <-0.5 & \\
\hline
\end{array}
$$
\end{table*}

The bolometric magnitudes were computed assuming a distance modulus for the LMC
of \( m-M=18.6 \).  The   \Teff\ values were obtained by means of the  
\Teff\ versus \( J-K \)\ and \Teff\ vs. \( V-K \)\ relations by BCP98, 
and these,  together with \Mbol\ and an assumed mass of 4.5 \Msun\, finally yield 
the gravity values. 
A small grid of synthetic spectra with various lithium abundances was
generated for the derived parameters. We found that the photometric 
\Teff\ did not allow a good fit of the spectra of the cooler stars.
Therefore, in order to derive the abundances, new values of \Teff\ 
were set by looking at the best overall fit over the 6350--7100
\AA\ region.
AGB-tip stars are known to be long-period variables, thus we expect differences
between the \Teff\ derived from photometry and the \Teff\ used for 
spectroscopy  (photometric and spectroscopic observations were not 
simultaneous). 

 Moreover, we do not expect  the above mentioned transformations to be very 
accurate for AGB-TP stars. Therefore, in the following discussion, we will 
stick to the spectroscopic \Teff\ values. We will use, for instance, the 
\Mbol\ values that can be derived using backwards the spectroscopic 
\Teff\ and the  above-mentioned calibrations.  
The results are listed in Table \ref{magn}.

\section{Comparison with theoretical models}
\subsection{\object{NGC 1866}}

Despite the rather large uncertainties on the derived Li abundances,
we have a number of interesting results. 

 Starting from the bottom of the AGB, we can conclude that the early 
AGB stars in \object{NGC 1866} \textit{do show} lithium,
although the derived abundance is very low (\( \log N(\mathrm{Li})\simeq 0 \)).
This is consistent with the fact that for a large progenitor mass,
we do not expect lithium to be destroyed in the  evolutionary phases
preceding AGB but only to be diluted, when the convective envelope sinks
into the star. Fig. \ref{fig1} shows, however, that the residual abundance of our 
models (for a 4 \Msun) is  \( \log N(\mathrm{Li})\sim 2.0 \), i.e. much
larger than the observed one. It has  to be noticed that these models have 
a solar-type initial abundance (\( \log N(\mathrm{Li})\simeq 3.3 \)), which by 
comparison
with the early AGB value,  implies a dilution  factor \( \sim 20 \). 
If the initial abundance were somewhat lower, say \( \log N(\mathrm{Li})\simeq 2.7 \), 
the diluted abundance would be accordingly scaled. That, however, cannot fully 
explain the discrepancy: it is highly plausible that an additional mixing mechanism 
providing further dilution is at work.
The occurrence of such a mechanism is well established, at least for
lower mass red giants, by the low \element[][12]{C}/\element[][13]{C} and low lithium
found on the red giant branch \citep[e.g.][]{charbonnel95}. We could also suggest
that the lower abundances are due to a mass loss larger than that included in
the \citet{ventura00} models.  Anyway, the models can still be used
for comparison, as the memory of the initial value is lost after the first
lithium dilution phase.

In spite of this quantitative discrepancy, the presence of lithium
in the early AGB stars proves  that in this
phase the star preserves some lithium, as expected from theory. The
models do not predict any lithium production here, so we regard this
result as a firm point: prior to the AGB, lithium is present but heavily
diluted, and its abundance in the envelope sets an upper limit
to  its content during the following evolutionary phases,
in absence of further production. 

The spectrum of  the cooler star N1866\#3 shows the TiO bands and  
a relatively strong lithium line. However, this star is not yet as luminous
as we would expect for lithium production by HBB. It has to be noticed
that because of the somewhat lower quality of the spectra, the uncertainties 
on the abundance are rather large (see Table \ref{magn}),
and we cannot safely assume for this star a lithium content larger than that 
of the other,  hotter, early AGB ones.

Looking  at the most luminous stars N1866\#1 and \#2, we find a totally
different situation:  star \#2 has an abundance definitely larger than that
of  the early AGB ones ($\log N(\mathrm{Li}) \simgt 1.5 \pm 0.5$), 
so that \textit{we conclude that we are witnessing lithium production}. 
No lithium line is
detected in the spectra of the other star. This yields an upper limit at 
$\log N(\mathrm{Li}< -0.5 $; on this basis we can certainly suggest that lithium in
this star has been destroyed.

The amount of lithium  manufactured in NGC1866\#2 is not particularly large,
but because of the error size, we must wait for high dispersion spectra to
set more stringent limits. The abundance is, however, in the range expected
from the models (Fig. \ref{fig1}), also taking into account that it varies with the
thermal pulse phase.

The absolute bolometric magnitude of this star is actually at the lower
boundary of what we expect for the occurrence of HBB: (\Mbol$=-6$),  and
that might be an additional reason for the relatively low log~N(Li) value.
Fig. \ref{fig6} shows the Li vs. \Mbol\ along the same 4 \Msun\
evolutionary track of Fig. \ref{fig1}. 
The points corresponding to the observed stars
are also plotted. We see that indeed the star\#2 in \object{NGC 1866} is at the phase
in which the residual lithium is completely destroyed, while the star \#1 is
in a phase in which it is manufactured by HBB. The agreement with the
theoretical models is quite satisfactory.  Things are however  less clear
for the cluster \object{NGC 2031}.

\subsection{\object{NGC 2031}}

The stars in this  cluster show a Li abundance behaviour similar to that in
\object{NGC 1866}, but both the quality of some spectra and the fact that this cluster
is much less populated than \object{NGC 1866} make the interpretation less stringent.

First of all we have , for the stars in early AGB  phase, only one reliable Li 
determination. This is for star (\#5), that clearly shows a lithium line
of similar  strength to those in the corresponding stars of \object{NGC 1866}. The
abundance analysis gives log~N(Li)\(=0.2 \pm 0.2 \).
Star (\#6) turned out to have a spectral type later than  that expected on
the basis of its photometry, a fact that casts doubts on its cluster
membership. For  star \#3, whose spectrum has no lithium line, we could only
get an upper limit. 

Concerning the later AGB objects, star \#1 seems to be very similar to  star \#1 in 
\object{NGC 1866}: it shows no lithium (abundance \( < \)1) and it is at about the right 
luminosity to be in the first phase of lithium destruction due to HBB.
The  analysis of star \#2, whose spectra have however a relatively lower S/N
ratio, provides a high lithium abundance (\( \sim 2.5\pm0 .5 \))
\textit{but, according to the models, this star is not luminous enough to be
in the HBB phase!} \citet{mould93} suggested, on the base their
photometry and the very red B$-$R ($>3$), that this star could be a very bright
carbon star: in this case its large lithium abundance would imply a
classification as J-subtype carbon star \citep{bouigue54}.   This
possibility  is ruled out, however, by the fact that its spectrum looks
oxygen-rich.

\subsection{\object{NGC 2214} and \object{NGC 2107}}
 
The two bottom spectra of Fig. \ref{fig4} show the two stars in the
clusters used as comparison terms. As shown in the figure, \object{NGC 2214} presents
a spectrum of intermediate type between the early and the later AGB stars. 
The abundance analysis provides log N(Li) $\simeq -0.5 \pm 0.3$. FMB90
attribute a spectral class M to this star, but with strong uncertainty. NGC
2107 \#1 is, on the other hand, cooler and shows clear features of a class M
star (FMB90 assign an M4 type). The TiO bands are visible but, as expected for a
star with a smaller mass than predicted for lithium production, the 6707.8
\AA\   line is absent.

\begin{figure}
\centerline{\resizebox{8.8cm}{!}{\rotatebox{0}{\includegraphics{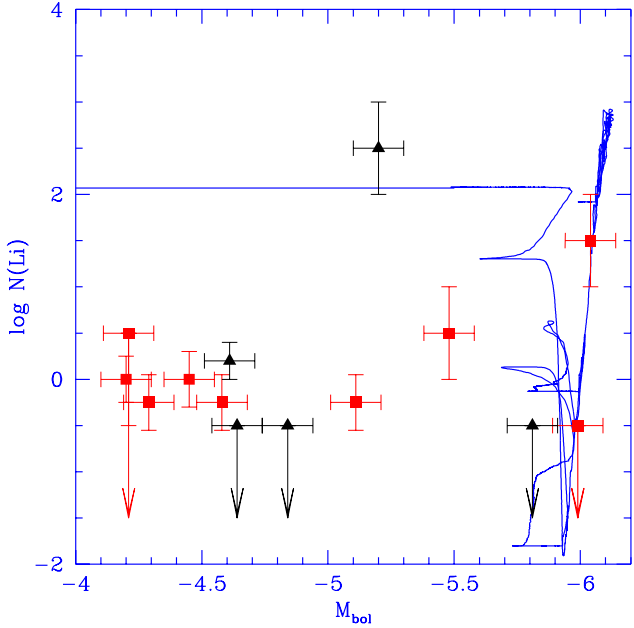}}}}
\caption[]{Lithium abundance versus \Mbol\ for the target stars  in NGC 1866 (squares)
and NGC 2031 (triangles). The  arrows represent upper limits. 
The  evolutionary track  of Fig. \ref{fig1} is also shown.
\label{fig6}
}
\end{figure}

\section{Evolutionary stage}

Although Fig. \ref{fig6} seems to provide a straightforward interpretation for
the \object{NGC 1866} lithium sequence, Fig. \ref{fig1} shows that another hypothesis
shall be tested, i.e. that the lithium poor star \#1 could actually be in a phase 
\textit{following} the Lithium production, when lithium is destroyed again.
In this case, however, the star  should have a larger bolometric luminosity and, 
moreover, should be surrounded by an extended circumstellar
envelope, easily detectable in the mid-infrared with ISOCAM. 
 In the ISOCAM field the three brightest sources in all bands coincide with the 
location of our stars \#1, \#2 and \#3, and their
fluxes in mJy are given in Table \ref{jansky}. The photometric results are
given in Table \ref{jansky}, where we have as well transformed the near-IR 
magnitudes, previously shown in Table \ref{magn} in Jy. 
The fluxes in the three bands at 4.5, 6.7 and 12\( \mu  \)\ show that
the most luminous source (and also reddest, if we consider  the ISOCAM color 
{[}4.5{]}$-${[}12{]})
is the star  which shows high lithium abundance. It is interesting to note that 
the brightest star in the K-band (\object{NGC 1866} \#1) 
is only the \textit{third} in the ISO bands, and is also the {}``bluest\char`\"{} in 
the ISO colors.  This  suggests a less evolved stage as an AGB star.
 
 In general the mid-IR colors of these stars are not {}``extreme\char`\"{}, indicating
that they are still in the initial stages of the AGB phase. This fact almost completely
discards the possibility of star \object{NGC 1866} \#1\ being in a later AGB phase,
when the lithium is destroyed again.

Another point to examine is the star distribution along the AGB of \object{NGC 1866}. 
The  model shown in  Fig. \ref{fig1} predicts a few more stars along the AGB: 
if we sample two stars in the first 40 \( \times 10^4 \)yr of evolution, 
we should find some other five in the following evolution, which lasts 
$\simeq 10^5$yr longer.
Most of these stars could be invisible in the optical and near infrared,
as they would be surrounded by a thick circumstellar envelope, but they should
be the brightest objects in the field in the mid-IR.
 Actually there are no  other bright
mid-IR sources in the ISOCAM field, apart from the three brightest sources
in the K-band under analysis in this paper. The absence of heavily obscured
AGB stars needs to be interpreted with caution, because of the small 
statistics, but it is telling us that possibly the duration of this 
heavily obscured AGB phase is shorter than that predicted by the models, probably
because of  stronger than expected mass loss rate experienced
by these massive AGB stars.

\begin{table}
\caption[]{Infrared fluxes for \object{NGC 1866} $^{\mathrm{a}}$}
\label{jansky}
$$
\begin{array}{p{0.15\linewidth}clllll}
\hline
ID      &  F_J  & F_H  & F_K  & 4.5\mu & 6.7\mu &  12\mu  \\
\noalign{\smallskip}
\hline
\noalign{\smallskip}
N1866\#1 &  62.5 & 99.2 & 80.4   &  6.5  &  3.7 &  1   \\
N1866\#2 &  57.5 & 89.7 & 70.7   &  8.4  &  5.0 & 2.6  \\
N1866\#3 &  40.3 & 53.0 & 49.0   &  7.2  &  4.7 & 1.7  \\
\noalign{\smallskip}
\hline
\end{array}
$$
\begin{list}{}{}
\item[$^{\mathrm{a}}$] Fluxes are in mJy
\end{list}
\end{table}

Although we do not have ISO data for the stars in \object{NGC 2031}, its luminous star
without lithium is at about the same absolute bolometric magnitude than the
similar star in \object{NGC 1866}, indicating that they probably represent a similar
evolutionary phase.

What is much less obvious is the evolutionary stage of the lithium rich star
in \object{NGC 2031} at \Mbol $\simeq -5.2$: this is not explained by HBB models at
all. Other lithium rich giants at similar magnitudes ar known, most of which
are  J stars. The so called `cool bottom processing' suggested by
\citet{wasserburg95} could be a way to explain this anomaly.
However, if the clusters \object{NGC 1866} and \object{NGC 2031} are indeed coeval, it is
intriguing that one cluster conforms to a `standard' mixing scheme, while
the other does not. It has to be noticed that, whatever the mixing process in AGB, lithium
is produced through the consumption of $^3$He. If this occurs at M$_{\rm
bol} \simeq -5$, no helium  is left for production at M$_{\rm bol} \simeq
-6$, where HBB becomes important. A last point to consider is that the
radial velocity of NGC2031\#2 markedly differs  from those of the other stars in NGC
2031, also taking into account the large errors. 
As this cluster field is very crowded, it is also possible that it is a
background star.

\section{Conclusions}

\begin{enumerate}
\item The early AGB stars in \object{NGC 1866} show lithium abundances which are 
roughly constant with increasing luminosity, consistent with the
lithium dilution expected to have taken place during the previous
evolutionary stages. The average abundance found is \( \log N(\mathrm{Li})\simeq
0.0\pm 0.5 \), implying, however, stronger than standard dilution.
This result is confirmed by at least  another early AGB star in \object{NGC 2031}, 
although the data derived from other stars analyzed in this cluster
are inconclusive;
\item We have detected three cool luminous AGB stars in \object{NGC 1866}. The
faintest one (\#3) shows a lithium abundance consistent with the remnant
abundance expected from an AGB star at the beginning of this phase as a
consequence of previous lithium dilution. The brightest one (\#1) does not
show any lithium and the upper limit derived  (\( \log N(\mathrm{Li})\leq -0.5
\)) suggests that this star plausibly is  in the phase of HBB preceding
lithium production (confirmed by the smaller mid-IR excess detected by
ISOCAM). The other most luminous AGB star (\#2) has a larger lithium
abundance (\( \log N(\mathrm{Li})\simeq 1.5\pm 0.5 \)), which we can attribute to
production by HBB.  On the basis of our models we would expect a few ($\sim
5$) other AGBs in the field of \object{NGC 1866}, while none were found, neither
luminous in the optical, nor in the mid-IR. This result put interesting
constraints on the duration of the AGB phase and the severity of the mass loss
during this phase.
\item The most luminous star in \object{NGC 2031} (\#1) is found to be similar to star \#1 
in \object{NGC 1866}. The second brightest star in the K-band in this cluster (\#2) 
has the largest lithium abundance in the sample, but Li production is not
predicted at its derived luminosity (HBB not active). 
Further observations are needed to confirm the abundance analysis and the
membership of this star.
\end{enumerate}
We conclude that, though many points still remain unclear in the interpretation 
of the observations,  we are in presence of an interesting sample
of stars, whose further careful analysis can  shed light on the
expected evolutionary paths. 
Observations at a higher dispersion are needed for these stars
to clarify their evolutionary status: for example, the presence of {\it s}-process 
elements in the spectra would be an important indicator of how many thermal
pulses the stars have gone through. Our best guess is that they still are at the
beginning of the TP phase, so that we should \textit{not} expect a sensible
{\it s}-process abundance enhancement.

\begin{acknowledgements}
CM, VT and FD acknowledge discussions with Paolo Ventura and Paolo Persi, and
the support of the Italian Ministry of University, Scientific
Research and Technology (MUIR) within the Cofin 1999 Project: ``Stellar
Dynamics and Stellar Evolution in Globular Clusters: a Challenge for the New
Astronomical Instruments". 
PGL acknowledges support from grant PB97-1435-C02-02 from
the Spanish Direcci\'on General de Ense\~nanza Superior e Investigaci\'on 
Cient\'\i fica (DGESIC).

\end{acknowledgements}

\bibliographystyle{aa}

\end{document}